\documentclass[twocolumn,prb,superscriptaddress]{revtex4}

\usepackage{graphicx}% Include figure files
\usepackage{dcolumn}% Align table columns on decimal point
\usepackage{bm}% bold math
\usepackage{amssymb, amsmath}
\usepackage{color}
\usepackage{slashed} 

\begin{document}
\title{Observation of quantum corrections to conductivity up to optical
  frequencies}

\author{P. Neilinger}
\affiliation{Department of Experimental
  Physics, Comenius University, SK-84248 Bratislava, Slovakia}

\author{J. Gregu\v{s}}
\affiliation{Department of Experimental
  Physics, Comenius University, SK-84248 Bratislava, Slovakia}

\author{D. Manca}
\affiliation{Department of Experimental
  Physics, Comenius University, SK-84248 Bratislava, Slovakia}

\author{B. Gran\v{c}i\v{c}}
\affiliation{Department of Experimental
  Physics, Comenius University, SK-84248 Bratislava, Slovakia}

\author{M.~Kop\v{c}\'{i}k}
\affiliation{Centre of Low Temperature Physics, Institute of
  Experimental Physics SAS, Watsonova 47, 040 01 Ko\v{s}ice, Slovakia}

\author{P.~Szab\'{o}}
\affiliation{Centre of Low Temperature Physics, Institute of
  Experimental Physics SAS, Watsonova 47, 040 01 Ko\v{s}ice, Slovakia}

\author{P.~Samuely}
\affiliation{Centre of Low Temperature Physics, Institute of
  Experimental Physics SAS, Watsonova 47, 040 01 Ko\v{s}ice, Slovakia}

\author{R. Hlubina}
\affiliation{Department of Experimental
  Physics, Comenius University, SK-84248 Bratislava, Slovakia}

\author{M.~Grajcar}
\affiliation{Department of Experimental
  Physics, Comenius University, SK-84248 Bratislava, Slovakia}
\affiliation{Institute of Physics, Slovak Academy of Sciences,
  D\'{u}bravsk\'{a} cesta, Bratislava, Slovakia}

\begin{abstract}
It is well known that conductivity of disordered metals is suppressed
in the limit of low frequencies and temperatures by quantum
corrections. Although predicted by theory to exist up to much higher
energies, such corrections have so far been experimentally proven only
for $\lesssim$80~meV. Here, by a combination of transport and optical
studies, we demonstrate that the quantum corrections are present in
strongly disordered conductor MoC up to at least $\sim$4~eV, thereby
extending the experimental window where such corrections were found by
a factor of 50. The knowledge of both, the real and imaginary parts of
conductivity, enables us to identify the microscopic parameters of the
conduction electron fluid.  We find that the conduction electron
density of strongly disordered MoC is surprisingly high and we argue
that this should be considered a generic property of metals on the
verge of disorder-induced localization transition.
\end{abstract}
\maketitle

At finite frequencies $\omega$, the optical conductivity of any
material is a complex quantity,
$\sigma(\omega)=\sigma^\prime(\omega)+i\sigma^{\prime\prime}(\omega)$.
In the limit of low frequencies, only the conduction band contributes
to the real part of the optical conductivity of a metal; the complex
conductivity of this band is customarily described by the Drude
formula\cite{Ziman72,Scheffler05}
\begin{equation}
\sigma(\omega)=\frac{\sigma_0}{1-i\omega/\Gamma},
\label{eq:drude}
\end{equation}
where $\Gamma=1/\tau$ is the relaxation rate determined by the
collision time of the electrons $\tau$. Within the simplest model of
free electrons the classical dc conductivity is given by
$\sigma_0=ne^2\tau/m$, where $n$ is the electron concentration and $m$
is the electron mass.\cite{Ziman72}

With increasing disorder strength, $\Gamma$ increases and $\sigma_0$
decreases, until at some critical disorder level the
(three-dimensional, 3D) metal turns into an insulator, as predicted
long ago by Anderson.\cite{Anderson58} At that point
$\sigma^\prime(0)$ vanishes in the limit of low temperatures $T$ and,
if Eq.~\eqref{eq:drude} were to apply, the optical conductivity
$\sigma(\omega)$ would vanish identically for all $\omega$. However,
this is unphysical, since at nonzero frequencies the absorption has to
be finite even in the insulating state.  Therefore, in the insulating
state, Eq.~\eqref{eq:drude} can not be valid. Instead, the
low-frequency conductivity $\sigma^\prime(\omega)$ has to grow with
$\omega$ and, by continuity, the same behaviour has to be expected
also on the metallic side of the metal-insulator transition.

It is in fact well known that, in {\it weakly disordered} 3D metals,
the conductivity in the small-$\omega$ and $T$ limit can be described
as $\sigma(\omega)=\sigma_{\rm reg}(\omega) +\delta\sigma(\omega)$,
where $\sigma_{\rm reg}(\omega)\approx \sigma_0$ is the regular part
of the conductivity and $\delta\sigma(\omega)$ is the quantum
correction which grows with $\omega$.  Two mechanisms have been
proposed for the latter: it can be either due to the so-called
weak-localization corrections,\cite{Gorkov79} or due to interaction
effects.\cite{Altshuler79} It is remarkable that, up to numerical
prefactors, at $T=0$ both mechanisms yield the same functional form of
the quantum correction.\cite{Altshuler85} Its real part,
$\delta\sigma^\prime(\omega)$, can be written in a unified way as
\begin{equation}
\delta\sigma^\prime(\omega)\approx
\mathcal{Q}^2\sigma_0\left(-1+\sqrt{\omega/\Gamma}\right).
\label{eq:corrections}
\end{equation}
The quantum correction is finite at $\omega\lesssim \Gamma$ and its
magnitude is characterized by a dimensionless number $\mathcal{Q}$, to
be called quantumness. We emphasize that our parametrization
Eq.~\eqref{eq:corrections} reflects the fact that the quantum
correction has to diminish the conductivity.

Numerical simulations by Weisse within the weak-localization scenario
at temperature $T=0$ have shown that, at not too high frequencies,
Eq.~\eqref{eq:corrections} is qualitatively valid not only for weakly
disordered metals, but in a {\it broad range of disorder strengths} in the
metallic phase.\cite{Weisse04} Throughout the metallic phase, Weisse's
data indicate that $\mathcal{Q}$ is of the same order of magnitude as
$\hbar\Gamma/\varepsilon_F$, where $\varepsilon_F$ is the Fermi
energy: in the limit of weak disorder this result is well
known;\cite{Altshuler85} in the opposite limit when the
metal-insulator transition is approached, the quantumness
$\mathcal{Q}\rightarrow 1$ and $\hbar\Gamma$ is comparable with
$\varepsilon_F$.\cite{note_weisse}

In order to generalize the above formulae for $\sigma^\prime(\omega)$
to finite temperatures, we follow the recipes of Fermi liquid theory
and replace $\omega$ by $\Omega=\sqrt{\omega^2+\gamma(T)^2}$
where $\gamma(T)$ is a temperature-dependent scattering rate which
depends on the mechanism for the quantum corrections: in case of
weak-localization corrections it is the phase-breaking scattering
rate, whereas for interactions effects $\gamma(T)=\pi k_B T/\hbar$.
Detailed calculations\cite{Altshuler85} for weakly disordered metals
do confirm such a procedure, again up to numerical prefactors.

Motivated by Weisse's data and analytical theory for weakly disordered
metals, we postulate (throughout the metallic phase) the following
simple formula for the frequency- and temperature-dependence of the
real part of the optical conductivity which features the Drude
behaviour Eq.~\eqref{eq:drude} in the high-frequency limit $\Omega\geq
\omega^\ast$ and, at the same time, for $\Omega<\omega^\ast$ takes
into account the 3D quantum correction Eq.~\eqref{eq:corrections}:
\begin{eqnarray}
\sigma^\prime(\omega,T)&=&\sigma_0
\left[1-\mathcal{Q}^2+\mathcal{Q}^2\sqrt{\Omega/\Gamma}\right],
\quad \text{if } \Omega<\omega^\ast,
\label{eq:comprehensive1}
\\
\sigma^\prime(\omega,T)&=&\frac{\sigma_0}{1+(\Omega/\Gamma)^2}, 
\qquad \hspace{1.65cm} \text{if } \Omega\geq \omega^\ast.
\label{eq:comprehensive2}
\end{eqnarray}
The model Eqs.~(\ref{eq:comprehensive1},\ref{eq:comprehensive2})
depends on three parameters: $\sigma_0$, $\mathcal{Q}$, and
$\Gamma$. Once these are known, the magnitude of the crossover
frequency $\omega^\ast$ follows from assuming that the function
$\sigma^\prime(\omega,T)$ is continuous.\cite{methods} We treat
$\mathcal{Q}$ and $\Gamma$ as independent, while for $\sigma_0$ we
keep the classical expression $\sigma_0=ne^2/(m\Gamma)$.

\begin{figure}
\centering
\includegraphics[width=9cm]{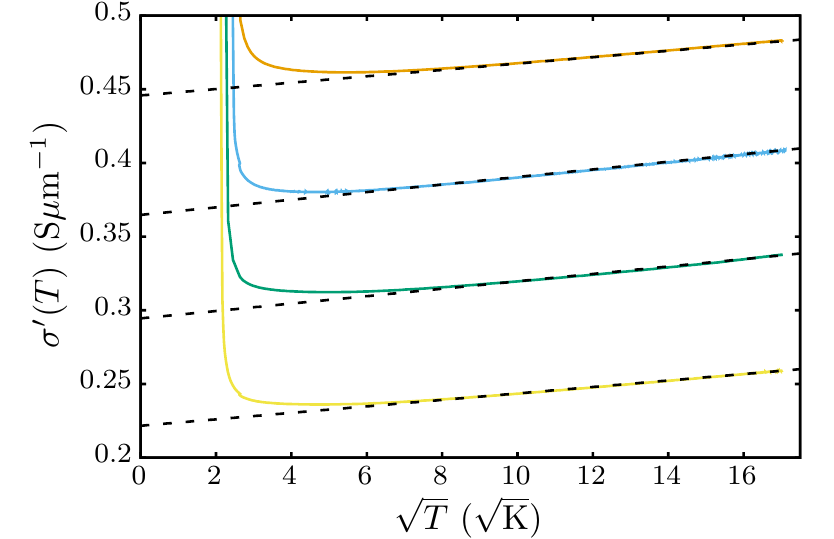}
\caption{Temperature dependence of dc conductivity for a series of
  5~nm thick MoC films. The sheet resistance of the films at room
  temperature is (from bottom to top) $R_\Box=720$,~590,~500, and
  415~$\Omega$. Fits to Eq.~\eqref{eq:comprehensive1} are shown as
  well.}
\label{fig:GT}
\end{figure}

The most comprehensive experimental test of the formula
Eq.~\eqref{eq:comprehensive1} has been performed on the amorphous
Nb:Si system close to its metal-insulator transition.\cite{Lee00} For
samples at the metallic side of the transition, very good agreement
has been found with a slightly generalized version of
Eq.~\eqref{eq:comprehensive1} for temperatures $T\lesssim 16$~K and at
frequencies up to 1~THz, both corresponding to $\hbar\Omega \lesssim
4$~meV. However, there are two reasons to expect that in very dirty
systems such as Nb:Si Eq.~\eqref{eq:comprehensive1} actually applies
in a much wider frequency range: First, close to the metal-insulator
transition $\hbar\Gamma$ should be comparable with the
bandwidth.\cite{Altshuler85} Second, the tunneling density of states
in amorphous Nb:Si is reduced at least up to 200~meV from the Fermi
level.\cite{Bishop85} This latter observation also suggests that the
quantum correction to conductivity in amorphous Nb:Si is due to
interaction effects.

The goal of the present paper is to look for quantum corrections to
conductivity in a much broader frequency range and to check whether
they can be observable even at optical frequencies. To this end, we
have chosen to study the highly disordered conductor
MoC,\cite{Lee90,Szabo2016} since in this material quantum corrections
have been observed in transport measurements up to~300~K,\cite{Lee94}
corresponding to $\hbar\Omega\approx 80$~meV.  

The amount of disorder in MoC can be conveniently tuned by varying the
Mo:C stoichiometry and/or by the film thickness: both, the reduction
of film thickness and the increase of the carbon content, lead to an
increase of the sheet resistance $R_\Box$.  In the present work, we
study two sets of samples: in the first set, we prepared films with
thickness $d$=5~nm and varying Mo:C stoichiometry, while in the second
set, we have studied films at fixed stoichiometry but with varying
thickness. Details on preparation of the MoC films are in the
supplementary material.\cite{methods}

The temperature dependence of dc conductivity $\sigma^\prime(T)$ for
the set of samples with fixed thickness is presented in
Fig.~\ref{fig:GT}.  In a transport measurement $\omega\approx 0$ and
therefore $\Omega=\gamma(T)$.  As can be seen, $\sigma^\prime(T)$
exhibits very good scaling with the square root of temperature from
$T_{\rm min}\approx~50$~K up to room temperature, precisely as
expected according to~Eq.~\eqref{eq:comprehensive1} in case of
dominant interaction effects with $\gamma(T)=\pi k_B T/\hbar$.  The
deviations from this scaling below $T_{\rm min}$ are caused by the
superconducting transition and the associated fluctuation
conductivity.  Moreover, dimensional crossover between 3D and 2D
quantum corrections is expected to occur at temperatures comparable
with $T_{\rm min}$.\cite{methods}

As can be seen from Fig.~\ref{fig:GT}, the two terms
in~Eq.~\eqref{eq:comprehensive1} exhibit quite different evolution
with stoichiometry: the extrapolated $\Omega=0$ value of the
conductivity $\sigma^\prime(0)$ decreases when the metal-insulator
transition is approached, whereas the coefficient in front of
$\sqrt{T}$ is roughly constant. Similar behaviour has been
observed previously in Nb:Si\cite{Bishop85,Lee00} and in
TiO$_x$;\cite{Vlekken91} it is also consistent with our model
Eqs.~(\ref{eq:comprehensive1},\ref{eq:comprehensive2}).\cite{methods}

\begin{figure}
\centering
\includegraphics[width=9cm]{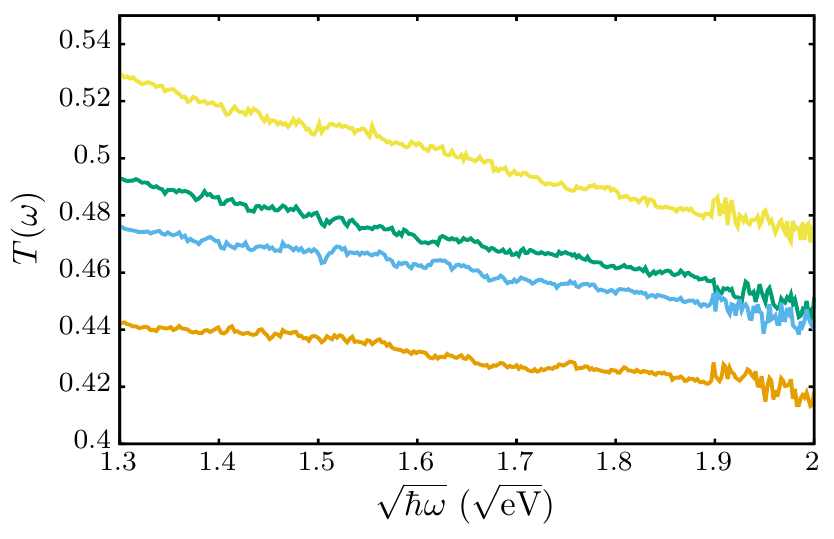}
\caption{Frequency dependence of transmission through MoC thin films
  on sapphire substrates for the same set of films and with the same
  colour coding as in Fig.~\ref{fig:GT}.}
\label{fig:Tr}
\end{figure}

In Fig.~\ref{fig:Tr} we show the optical transmission ${\cal
  T}(\omega)$ in a broad frequency range for the same set of films on
sapphire substrates as in Fig.~\ref{fig:GT}.  The absence of any
spectral features indicates that interband transitions are absent in
this range, a point we will come to later.  Similar featureless
transmission data is also obtained for the set of films with varying
thickness. This indicates that the details of microstructure are not
important for the phenomena we observe and that, for both sets of
samples, the crucial control parameter is the degree of
disorder.\cite{methods}

\begin{figure*}
\centering
\includegraphics[width=13cm]{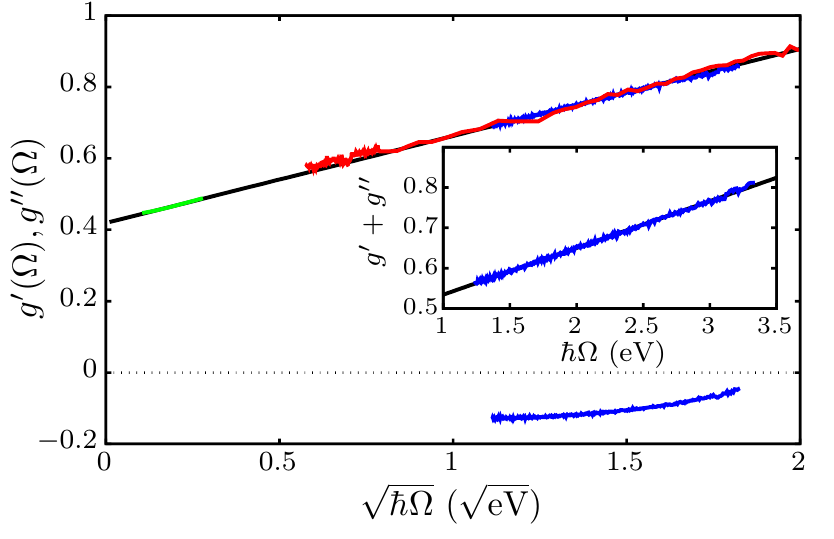}
\caption{Dimensionless sheet conductance $g(\Omega)$ of a 5~nm thick
  MoC film with room temperature sheet resistance
  $R_\Box=720~\Omega$. Green: data obtained from the temperature
  dependence of the dc conductivity in Fig.~\ref{fig:GT} assuming
  $\gamma(T)=\pi k_B T/\hbar$. Red: data from optical
  transmission. Blue: real (positive) and imaginary (negative) parts
  of $g(\Omega)$ determined by ellipsometry. Black line: fit of the
  real part to~Eq.~\eqref{eq:comprehensive1}. The inset shows that the
  anomalous terms proportional to $\sqrt{\Omega}$ perfectly cancel in
  $g^\prime(\Omega)+g^{\prime\prime}(\Omega)$.}
\label{fig:cond_exp}
\end{figure*}

Assuming $|\sigma^{\prime\prime}(\omega)|\ll \sigma^\prime(\omega)$,
the real part of the dimensionless sheet
conductance\cite{note_conductance} of the MoC films
$g(\omega)=Z_0\sigma(\omega)d$, where $Z_0$ is the impedance of free
space, can be calculated from the transmission ${\cal
  T}(\omega)$.\cite{methods} The thus obtained conductance
$g^\prime(\omega)$ of a MoC film with thickness $d$=5 nm and room
temperature sheet resistance $R_\Box=720~\Omega$ is shown in
Fig.~\ref{fig:cond_exp}. Also shown in Fig.~\ref{fig:cond_exp} is the
ellipsometry data for both components of $g(\omega)$ which is obtained
in a somewhat more narrow frequency range. Since at optical
frequencies $\omega\gg \gamma(T)$, we do not have to distinguish
between $\omega$ and $\Omega$.  The temperature dependence of the
conductivity for a sample from the same batch is replotted here from
Fig.~\ref{fig:GT} as well, assuming $\gamma(T)=\pi k_B T/\hbar$.  Note
the very good agreement between all three data sets.

Since conductivity $\sigma$ and dimensionless conductance $g$ differ
only by a multiplicative constant, when talking about frequency- and
temperature dependence, from now on we will use these terms
interchangeably.

The data presented in Fig.~\ref{fig:cond_exp} is the main result of
this paper. It shows that the real part of conductivity of strongly
disordered MoC thin films is very well described
by~Eq.~\eqref{eq:comprehensive1} in a broad range of frequencies from
$\hbar\Omega\approx 14$~meV up to at least $\hbar\Omega\approx 4$~eV.
Although this is expected from the theoretical point of view since the
scattering rate $\Gamma$ in dirty metals close to the metal-insulator
transition is huge, to our knowledge, until now it has not been
demonstrated experimentally.  As a consequence, this fact is not
generally adopted and it is often incorrectly assumed that quantum
corrections are not present at room temperature or at optical
frequencies. 

Equally remarkable are the results for the imaginary part of
conductivity $\sigma^{\prime\prime}(\omega)$ which are also presented
in Fig.~\ref{fig:cond_exp}. It should be pointed out that, unlike the
real part of conductivity which is (in the studied frequency range)
determined only by the contribution of the conduction band, there
exists an additional contribution to $\sigma^{\prime\prime}(\omega)$
from the bound electrons, $\sigma^{\prime\prime}_{\rm
  bound}(\omega)=-\epsilon_0(\epsilon_\infty-1)\omega$ where
$\epsilon_0$ is the permittivity of vacuum and $\epsilon_\infty$ is
the bound-electron contribution to the static dielectric
constant. Thus the presence of a negative contribution to
$\sigma^{\prime\prime}(\omega)$ is by itself not surprising. However,
the experimental data in Fig.~\ref{fig:cond_exp} clearly indicate that
$\sigma^{\prime\prime}(\omega)$ is not linear in frequency.  In fact,
the data contain an anomalous term
$-\mathcal{Q}^2\sigma_0\sqrt{\omega}$ with the same magnitude and
opposite sign as in the real part Eq.~\eqref{eq:comprehensive1}.  Such
a term is required to be present by the Kramers-Kronig relations and
in the inset to Fig.~\ref{fig:cond_exp} we demonstrate that, as was to
be expected, the anomalous terms perfectly cancel in the sum
$\sigma^{\prime}(\omega)+\sigma^{\prime\prime}(\omega)$.

The next natural question to ask is: what are the values of the
parameters $\sigma_0$, $\mathcal{Q}$, and $\Gamma$ which enter
Eq.~(\ref{eq:comprehensive1})?  From Fig.~\ref{fig:cond_exp} we have
access to only two parameters: the $\Omega=0$ value of the
conductivity $\sigma^\prime(0)$ and the coefficient in front of
$\sqrt{\Omega}$.  On the other hand, if we could extend our
measurements to higher frequencies and measure the crossover scale
$\omega^\ast$ predicted by
Eqs.~(\ref{eq:comprehensive1},\ref{eq:comprehensive2}), this would
give us the needed third data point. Unfortunately, at frequencies
above $\hbar\omega\sim 4$~eV the transmission of our sapphire
substrates is influenced by impurity absorption and therefore we can
not measure $\omega^\ast$ directly.

Nevertheless, one can approximately determine the complex conductivity
in the whole frequency range from a measurement of both,
$\sigma^{\prime}(\omega)$ and $\sigma^{\prime\prime}(\omega)$, in a
finite interval of frequencies.\cite{Dienstfrey2001} The key
observation is that the real and imaginary parts of conductivity have
to satisfy the Kramers-Kronig relations, and therefore one can write
down integral equations for the unknown conductivity outside the
measured range. However, since such analytic continuation problem is
ill-posed, additional simplifying assumptions have to be made. We have
used the following procedure for the prolongation of the
$\sigma^\prime(\omega)$ data, which turned out to be quite robust: We
start by choosing a value of $\Gamma$. Having made this choice, we can
unambiguously find $\sigma_0$ and $\mathcal{Q}$ from fitting the real
part of conductivity to Eq.~(\ref{eq:comprehensive1}). With known
$\sigma_0$, $\mathcal{Q}$, and $\Gamma$, the real part of conductivity
is known from Eqs.~(\ref{eq:comprehensive1},\ref{eq:comprehensive2})
on the entire real axis. Next we calculate, making use of the
Kramers-Kronig relations, the imaginary part of the conductivity
$\sigma^{\prime\prime}(\omega)$. Finally we adjust the value of
$\Gamma$ so that good agreement with $\sigma^{\prime\prime}(\omega)$
of the conduction band is obtained. Note that in order to obtain the
latter, the contribution $\sigma^{\prime\prime}_{\rm bound}(\omega)$
of the bound electrons should be subtracted from the ellipsometric
data for the imaginary part of conductivity in
Fig.~\ref{fig:cond_exp}.

\begin{figure}[t]
\centering
\includegraphics[width=9cm]{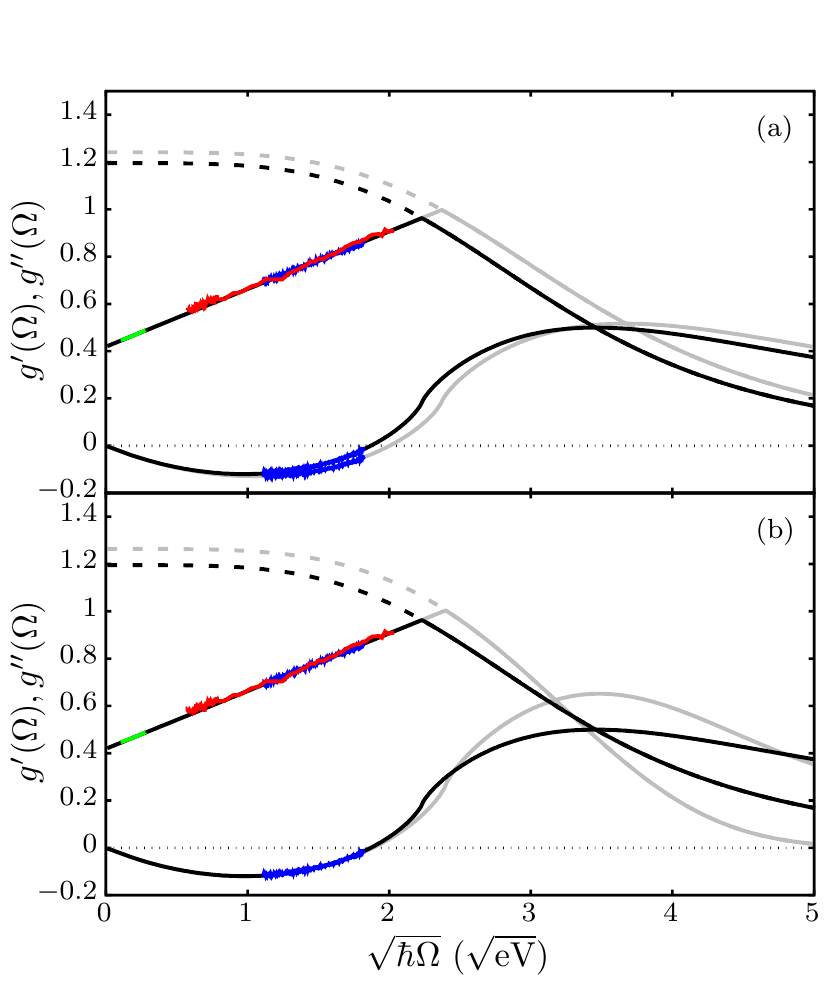}
\caption{\label{fig:Gr} Results of the prolongation procedure
  described in the text applied to the data presented in
  Fig.~\ref{fig:cond_exp}.  The colour coding of the experimental data
  is the same as in Fig.~\ref{fig:cond_exp}.  The dashed lines denote
  the continuation of the high-frequency behaviour (i.e. conductance
  in absence of quantum corrections) to low frequencies. (a) The real
  part of conductivity is modelled by
  Eqs.~(\ref{eq:comprehensive1},\ref{eq:comprehensive2}) for
  $\epsilon_\infty=1$ (gray line) and $\epsilon_\infty=1.4$ (black
  line).  (b) Assuming $\epsilon_\infty=1.4$, the high-frequency form
  of the real part is modelled by Eq.~(\ref{eq:comprehensive2}) (black
  line) and by a Gaussian described in the text (gray line).}
\label{fig:inversion}
\end{figure}

The results of the described prolongation procedure are presented in
Fig.~\ref{fig:inversion} and the corresponding fitting parameters are
summarized in Table~\ref{tab:fit}. One observes that all presented
prolongations of $\sigma^{\prime}(\omega)$ fit the
$\sigma^{\prime\prime}(\omega)$ of the conduction band very well. The
extracted values of the parameters $\sigma_0$, $\mathcal{Q}$, and
$\Gamma$ do depend on the assumed value of $\epsilon_\infty$, but
the variation between the results for $\epsilon_\infty=1$ and
$\epsilon_\infty=1.4$ is at most 10~\%. It can be shown that the
most likely value of $\epsilon_\infty$ is bounded by these two
values.\cite{methods}

In order to further test the robustness of our prolongation procedure,
we have modelled the real part of conductivity at high frequencies
$\omega\geq \omega^\ast$ by the Gaussian formula
$\sigma^\prime(\omega)=\sigma_0\exp(-\omega^2/\Gamma^2)$ instead of
Eq.~(\ref{eq:comprehensive2}). As shown in Fig.~\ref{fig:inversion},
the resulting fits of $\sigma^{\prime\prime}(\omega)$ are equally good
as when Eq.~(\ref{eq:comprehensive2}) is used, and the scatter in the
parameters $\sigma_0$, $\mathcal{Q}$, and $\Gamma$ is at the
level of 15~\% or less. We have also checked that when the cusp at
$\omega=\omega^\ast$ in the frequency dependence of
$\sigma^\prime(\omega)$ is smoothened, the results of the prolongation
procedure exhibit only marginal changes.\cite{methods} Based on all of
this evidence we conclude that our prolongation procedure is robust
and the extracted parameters $\sigma_0$, $\mathcal{Q}$, and $\Gamma$
are known with $\lesssim 15$~\% uncertainty.

According to Table~\ref{tab:fit}, the quantum corrections are
appreciable, $\mathcal{Q}\approx 0.81$. Therefore the $T=0$ limit of
the dc conductivity, $\sigma^\prime(0)=(1-\mathcal{Q}^2)\sigma_0$, is
reduced from the classical value $\sigma_0$ roughly by a factor of
3. If one were to interpret the $T=0$ limit of the measured dc
conductivity data of the studied sample as $\sigma_0$ (as is
frequently done), one would overestimate the scattering rate $\Gamma$
by the same factor of 3.

\begin{table}[t]
\centering
\begin{tabular}{ |p{2cm}||p{1cm}|p{1cm}||p{1cm}| p{1cm}| }
	\hline
	\ & \multicolumn{2}{|c||}{Drude, Eq.~\eqref{eq:comprehensive2}} 
          & \multicolumn{2}{|c|}{Gaussian}\\
	\hline
	$\epsilon_\infty$  & 1.0 & 1.4 & 1.0 & 1.4 \\
	\hline
	\hline
	$g_0=Z_0\sigma_0 d$   &  1.25  &  1.20 &  1.32  & 1.26  \\
	$\mathcal{Q}^2$ &  0.66  &  0.65 &  0.68 & 0.67  \\
%	$\Gamma/2\pi $ (THz)    &  2780 &  2450 &  3280  & 2900  \\
	$\hbar\Gamma$  (eV)  & 11.5 &  10.1 &  13.6  & 12.0  \\
	\hline
	\hline
	$n$ ($10^{23}$ cm$^{-3}$)     
&  4.1 &  3.5 &  5.1 &  4.3\\
	\hline
\end{tabular}
\caption{Parameters obtained from the fits of the sheet conductance
  presented in Fig.~\ref{fig:inversion}. Two models of the
  high-frequency form of the real part of conductance were used
  (Drude, Eq.~\ref{eq:comprehensive2}, and Gaussian described in the
  text) and two values of the dielectric costant were assumed,
  $\epsilon_\infty=1.0$ and $\epsilon_\infty=1.4$. The
  conduction electron density $n$ was estimated from $\sigma_0$ and
  $\Gamma$ making use of $\sigma_0=ne^2/(m\Gamma)$, where for $m$ we
  take the free electron mass. }
\label{tab:fit}
\end{table}

Regarding the energy scale $\hbar\Gamma$, it is surprisingly large,
$11.85\pm 1.75$~eV. This does make sense, however: up to 4~eV,
the real part of conductivity is described by
Eq.~(\ref{eq:comprehensive1}) without any noticeable higher powers of
frequency. This must mean that the crossover scale $\hbar\omega^\ast$
is by a wide margin larger than 4~eV. If one further observes that for
$\mathcal{Q}\approx 0.81$ we have $\hbar\omega^\ast\approx
0.5\hbar\Gamma$,\cite{methods} the large value of $\hbar\Gamma$ seems
to be inevitable.

What is even more surprising is that the electron concentration in the
conduction band is very large, more than twice as large as in metallic
aluminum.  We believe that this is a consequence of the large value of
$\Gamma$: the individual electronic bands which are separated by
energy less than $\hbar\Gamma$ lose their identity and merge
together. In order to estimate the electron concentration predicted by
such a picture, let us start by considering cubic MoC which
crystallizes in the rocksalt structure with lattice constant
4.27~\AA\cite{Krasnenko12} and concentration of one type of atoms
$n_{\rm at}=5.1\times 10^{22}$~cm$^{-3}$. The valence electron
configurations of the Mo and C atoms are 4d$^5$ 5s$^1$ and 2s$^2$
2p$^2$, respectively.  According to band-structure
calculations,\cite{Kavitha16,Krasnenko12} the relevant 4d and 5s
states of molybdenum as well as the 2s and 2p states of carbon are
within $\pm\hbar\Gamma$ from the Fermi energy. Therefore it is
reasonable to assume that the conduction electron fluid is formed by
all 10 valence electrons and the corresponding electron density is
$n=10\times n_{\rm at}=5.1\times 10^{23}$~cm$^{-3}$, a value within
the error bar of the data in Table~\ref{tab:fit}. As a matter of fact,
the value of $n_{\rm at}$ in a highly disordered material is actually
expected to be lower than the value for a perfect crystal which we
have used; this (as well as an excess of carbon atoms which we also
did not take into account) should decrease our estimate of $n$ and
improve the agreement with Table~\ref{tab:fit}. It is also worth
pointing out that the absence of interband transitions up to~4~eV (see
Fig.~\ref{fig:Tr}) provides an additional non-trivial consistency
check of our proposal that the conduction electron ``band'' is very
broad.

Because of the mechanism just described, we believe that the electron
concentration $n$ in the conduction fluid of a dirty metal close to
the metal-insulator transition should be (somewhat paradoxically)
generically large. In fact, e.g. in dirty NbN samples large values of
$n$ have already been observed: a naive analysis of the Hall
coefficient at $T=$20~K in films with resistivity $\rho\approx
100\:\mu\Omega$cm yields\cite{Destraz17} $n\approx 4.2\times
10^{23}$~cm$^{-3}$. However, quantum corrections to the Hall
coefficient are known to be present in similar samples of
NbN,\cite{Chand09} and therefore the quoted value of $n$ should be
taken as a lower bound to the actual electron concentration in the
conduction fluid. NbN crystallizes in the rocksalt structure with
lattice constant 4.39~\AA\cite{Shiino10} and concentration of one type
of atoms $n_{\rm at}=4.7\times 10^{22}$~cm$^{-3}$. It has the same
electron count of 10 valence electrons as MoC, therefore within our
picture we should expect $n=10\times n_{\rm at}=4.7\times
10^{23}$~cm$^{-3}$, which is in reasonable agreement with the Hall
estimate.

It is worth pointing out that the large values of electron
concentration $n$ reported in Table~\ref{tab:fit} imply a large Fermi
energy $\varepsilon_F$. In fact, making use of the free-electron
estimate $\varepsilon_F=\hbar^2\left(3\pi^2 n\right)^{2/3}/(2m)$ we
find $\varepsilon_F=20.65\pm 2.55$~eV. Nevertheless, the customary
parameter $k_F\ell$ characterizing the disorder level, defined by
$k_F\ell=2\varepsilon_F/(\hbar\Gamma)$, is quite small,
$k_F\ell\approx 3.5\pm0.1$, and this is consistent with the large
quantumness $\mathcal{Q}\approx 0.81$.

In conclusion, we have demonstrated that, in strongly disordered
metals on the verge of disorder-induced localization transition,
quantum corrections to conductivity may be present up to optical
frequencies.  This effect should be universal; therefore quantum
corrections should be added to the list of known
reasons\cite{Scheffler05} why the canonical Drude formula for the
frequency dependence of conductivity is hard to observe.  We speculate
that quantum corrections at infrared frequencies and above may already
have been observed previously, but they were interpreted in a
different way. Most notable candidates are liquid
mercury\cite{Inagaki81} and perhaps also Si:P.\cite{Gaymann95} It
therefore seems worthwhile to also take the quantum corrections into
account in models used for spectroscopic ellipsometry.

We have likewise demonstrated how the combined knowledge of both, the
real and imaginary parts of the optical conductivity, can be used to
extract microscopic parameters of the conduction electron fluid in
dirty metals which are not directly accessible otherwise - such as the
quantumness $\mathcal{Q}$, the scattering rate $\Gamma$, and
especially the electron concentration $n$. We have found that $n$ is
very large in MoC close to the metal-insulator transition; its value
indicates that the conduction electron fluid is formed by all valence
electrons.  We have argued that the large value of $n$ should be a
generic property of dirty metals, since individual electronic bands
which are separated by energy less than $\hbar\Gamma$ lose their
identity and merge together.

\section*{Supplementary material}
\subsection*{MoC: sample preparation and characterization}
The MoC thin films were prepared by means of reactive magnetron
deposition from a Mo target in argon-acetylene atmosphere (both gases
used of purity 5.0) on c-cut sapphire wafers heated to 200 degrees
Celsius.  The flow rate of argon was kept fixed, whereas the flow rate
of acetylene was varied between depositions, in order to tune the Mo:C
stoichiometry.  During deposition, the magnetron current was held
constant at 200 mA implying a deposition rate $\approx 11$~nm/min. The
deposition time, and thus the thickness, was regulated by means of a
programmable shutter control interface with precision of 1~s. The
chamber was evacuated to approximately $5\times 10^{-5}$~Pa. More
details on preparation of the MoC films and their characterization
have been published before.\cite{Trgala14}

STM and STS measurements show that films with a thickness larger than
3~nm are spatially homogeneous, with typical rms roughness of
0.6~nm.\cite{Szabo2016}

Two sets of samples were studied: in the first set, we prepared films
with a constant thickness $d=5$ nm and the Mo:C stoichiometry was
changed by varying the flow rate of acetylene. The sheet resistances
of the samples at room temperature were $R_\Box= 415$, 500, 590, and,
720~$\Omega$.

In the second set, the stoichiometry was kept fixed at a value which
maximizes the superconducting $T_c$ for films with 30~nm thickness.
We have prepared samples with thickness $d=20$, 15, 10, and 5~nm with
sheet resistances $R_\Box= 96$, 120, 216, and 495~$\Omega$,
respectively.

\begin{figure}
\centering
\includegraphics[width=9cm]{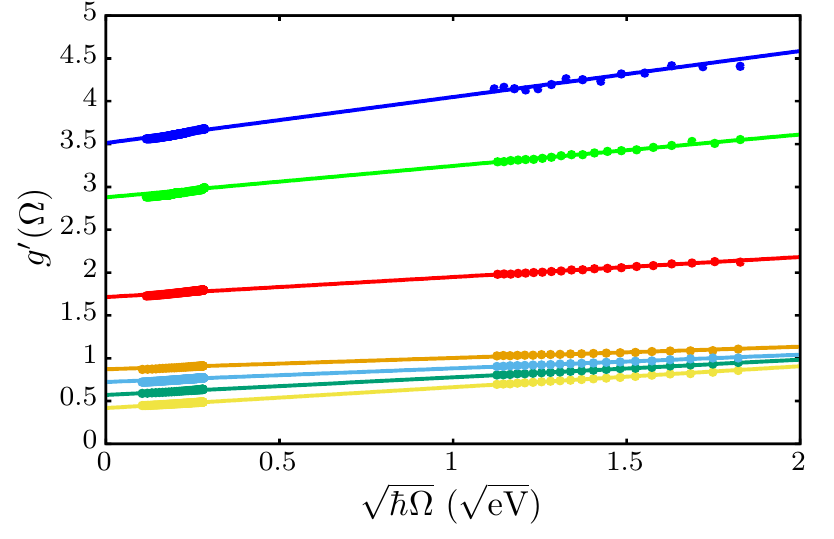}
\caption{Normalized sheet conductance for both sets of MoC thin
  films. The low-$\Omega$ and high-$\Omega$ data comes from the dc
  transport measurements assuming $\gamma(T)=\pi k_B T/\hbar$ and from
  ellipsometry, respectively. From top to bottom, the curves
  correspond to samples from the second set with sheet resistances
  $R_\Box= 96$, 120, and 216~$\Omega$, respectively.  The four curves
  at the bottom correspond to samples from the first set; the colour
  coding is the same as in Fig.~\ref{fig:GT}. Data for the film from
  the second set with $R_\Box= 495$~$\Omega$ is not plotted, since it
  is essentially the same as that for the film from the first set with
  $R_\Box=500$~$\Omega$. }
\label{fig:thickness}
\end{figure}

In Fig.~\ref{fig:thickness} we show the normalized sheet conductance
$g^\prime(\Omega)$ for both sets of MoC thin films.  The plotted data
is obtained from the temperature dependence of the dc conductivity and
from ellipsometry in the same way as in Fig.~\ref{fig:cond_exp}. Good
scaling with $\sqrt{\Omega}$ is found for all samples. The larger
sample-to-sample variations in the second set are a trivial
consequence of the varying sample thickness.

\subsection*{Optical measurements}
The frequency dependence of the real and imaginary part of sheet
conductance was measured by spectroscopic
ellipsometry\cite{Humlicek05} at room temperature in the wavelength
range 370 to 1000 nm using a rotating - compensator instrument
J. A. Woollam, M-2000V.

The transmission measurements were performed at normal incidence of
light to the sample surface.  In the visible and ultraviolet
frequency range we have used the UV VIS Carl Zeiss spectrometer and
the Ocean Optics spectrometer (USB650UV).  The transmission spectra in
infrared up to 2.5 $\mu$m were obtained by using the SPM2 Carl Zeiss
grating monochromator (gratings No456039 and No465645) with Hamamatsu
K1713-01 Si photodiode and PbS cell. At larger wavelengths up to 6
$\mu$m, a modified UR20 Carl Zeiss IR spectrometer with a LiF prism
was used.

The real part of the normalized sheet conductance $g^\prime(\omega)$
was extracted from the transmission data as follows.  Let us consider
transmission of light (at normal incidence) across a thin film with
complex normalized sheet conductance
$g(\omega)=g^\prime(\omega)+ig^{\prime\prime}(\omega)$ and thickness
$d$ which is deposited on a thick dielectric substrate (in our case
the substrate thickness is $\approx 430\:\mu$m) with a
purely real index of refraction $n_s(\omega)$. Let the transmission of
light from vacuum across the thin film into the substrate be
$\mathcal{T}_t(\omega)$ and let the corresponding reflection be
$\mathcal{R}_t(\omega)$. The transmission and reflection through the
interface substrate-vacuum are given by
$\mathcal{T}_s(\omega)=4n_s/(1+n_s)^2$ and
$\mathcal{R}_s(\omega)=1-\mathcal{T}_s(\omega)$, respectively, and the
total transmission through the system 'film + substrate' is
$$
\mathcal{T}(\omega)=\frac{\mathcal{T}_s(\omega)\mathcal{T}_t(\omega)}
{1-\mathcal{R}_s(\omega)\mathcal{R}_t(\omega)}.
$$
If the thickness $d$ satisfies the
constraints $d\ll c/\omega$ and $d\ll c|g(\omega)|/\omega$, then 
we can approximately write
\begin{eqnarray*}
\frac{\mathcal{R}_t(\omega)}{\mathcal{R}_s(\omega)}&\approx&
\frac{[1+g^\prime(\omega)/(n_s-1)]^2+[g^{\prime\prime}(\omega)/(n_s-1)]^2}
{[1+g^\prime(\omega)/(n_s+1)]^2+[g^{\prime\prime}(\omega)/(n_s+1)]^2},
\\ 
\mathcal{T}_t(\omega)&\approx&
\frac{\mathcal{T}_s(\omega)}
{[1+g^\prime(\omega)/(n_s+1)]^2+[g^{\prime\prime}(\omega)/(n_s+1)]^2}.
\end{eqnarray*}

For $n_s(\omega)$ we have used a three-term dispersion
formula\cite{Malitson62} according to which $n_s$ varies between 1.75
and 1.79 throughout the visible range. Assuming furthermore that
$g^\prime,|g^{\prime\prime}|\lesssim 1$ (see Fig.~\ref{fig:cond_exp})
one can show that $\mathcal{R}_s(\omega)\mathcal{R}_t(\omega)\ll 1$
and therefore the measured transmission is given by
$$
\mathcal{T}(\omega)\approx
\frac{\mathcal{T}_s(\omega)^2}{[1+g^\prime(\omega)/(n_s+1)]^2
+[g^{\prime\prime}(\omega)/(n_s+1)]^2}.
$$ 
When analyzing the experimental
data, we have neglected the $g^{\prime\prime}(\omega)$ term in the
denominator. Comparing with Fig.~\ref{fig:cond_exp} one can check a
posteriori that this procedure is justified.

\subsection*{Notes on the model 
Eqs.~(\ref{eq:comprehensive1},\ref{eq:comprehensive2})} The crossover
frequency $\omega^\ast$ of the model
Eqs.~(\ref{eq:comprehensive1},\ref{eq:comprehensive2}) is given by
$\Gamma$ with a $\mathcal{Q}$-dependent prefactor, $\omega^\ast=\Gamma
f(\mathcal{Q})$. The function $f(\mathcal{Q})$ is monotonic in the
whole interval between 0 and 1. In the limit of weak quantum
corrections, $\mathcal{Q}\ll 1$, it reduces to $f(\mathcal{Q})\approx
\mathcal{Q}$.  This is consistent with the expectation that
$\omega^\ast/\Gamma$ should vanish when $\mathcal{Q}\rightarrow 0$. In
the limit of strong quantum corrections $f(1)\approx 0.57$ and the
crossover frequency $\omega^\ast$ is a substantial fraction of
$\Gamma$, again in agreement with expectations.

Weisse's data indicate that $\omega^\ast$ stays finite when the
metal-insulator transition is approached,\cite{Weisse04} i.e. for
$\mathcal{Q}\rightarrow 1$, and from here it follows that $\Gamma$
also stays finite in this limit, as claimed in the main
text. Consequently, close to the metal-insulator transition, the
prefactor of the $\sqrt{\Omega}$ term in Eq.~\eqref{eq:comprehensive1}
varies only weakly with the disorder strength, whereas the
frequency-independent term $\sigma^\prime(0)$ drops to zero as the
transition is approached.

\subsection*{Estimate of the dielectric constant $\epsilon_\infty$}
By definition, $\epsilon_\infty=1-\lim_{\omega\rightarrow
  0}\sigma_{\rm bound}^{\prime\prime}(\omega)/(\epsilon_0\omega)$,
where $\sigma_{\rm bound}(\omega)$ is the interband
conductivity.\cite{note_epsilon} Making use of the Kramers-Kronig
relations we thus find
$$
\epsilon_\infty=1+\frac{2}{\pi\epsilon_0}
\int_0^\infty \frac{d\nu \sigma_{\rm bound}^\prime(\nu)}{\nu^2}
\approx 1+\sum_j\frac{\Omega_j^2}{\omega_j^2},
$$ 
where in the approximate equality we have modelled the interband
conductivity by a set of oscillators with frequency $\omega_j$ and
oscillator weights $\Omega_j^2$. From the f-sum rule we know that
$\sum_j \Omega_j^2=n_{\rm tot}e^2/(m\epsilon_0)-{\cal D}$, where
$n_{\rm tot}$ is the total electron density in the material and ${\cal
  D}$ is the oscillator weight of the conduction electrons.

As a rough estimate we will assume that, when the effect of a large
$\hbar\Gamma$ is taken into account, the spectrum of the material
consists of deep atomic-like electron levels labelled by $k$, and of a
single conduction band extending to infinity. In such case all
interband transitions $j$ take place between $k$ and the conduction
band, so that we can label them as $k$.  Moreover, we will assume that
$\Omega_k^2=Z_kn_{\rm at}e^2/(m\epsilon_0)$ where $Z_k$ is the number
of electrons in the atomic level $k$, whereby we approximately satisfy
the f-sum rule. Thus we arrive at the following estimate of
$\epsilon_\infty$:
\begin{equation}
\epsilon_\infty\approx 1+\sum_k\frac{Z_k\Omega^2}{\omega_k^2},
\qquad
\Omega^2=\frac{n_{\rm at}e^2}{m\epsilon_0},
\label{eq:epsilon}
\end{equation}
where (as an upper bound on $\epsilon_\infty$), for $\hbar\omega_k$ we
take the distance between the Fermi level and the atomic level $k$.

\begin{figure}
\centering
\includegraphics[width=9cm]{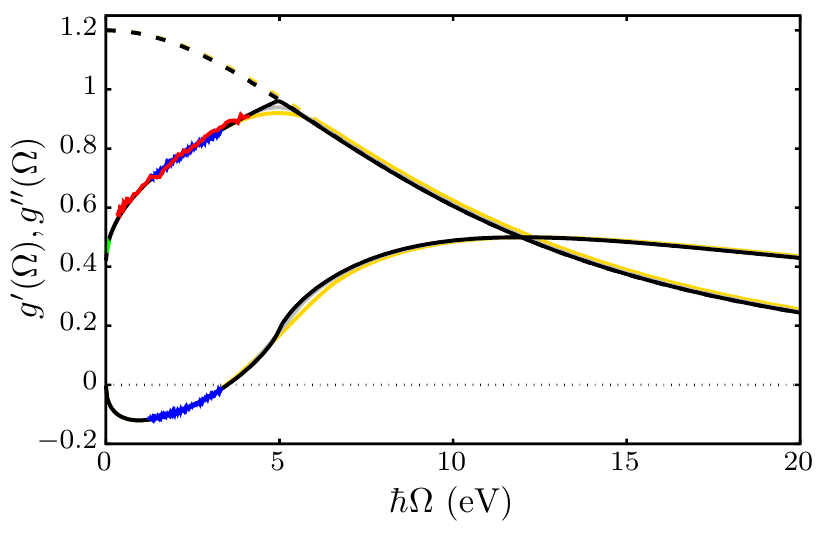}
\caption{Fits of the normalized sheet conductance data from
  Fig.~\ref{fig:cond_exp} by a smoothened Lorentzian model assuming
  $\epsilon_\infty=1.4$.  The colour coding of the experimental data
  is the same as in Fig.~\ref{fig:cond_exp}.  Note that a linear
  frequency scale is used. In the three sets of curves which are
  plotted we have used $\hbar\omega_{-}=3.3$, 4.1, and 5.0~eV, which
  correspond to the end of the ellipsometric data (gold), the end of
  the transmission data (silver), and a frequency close to $\omega^\ast$
  (bronze), respectively. The dashed lines denote the continuation of
  the high-frequency Drude formula to low frequencies for all three
  values of $\omega_{-}$.}
\label{fig:splines}
\end{figure}

As a test of our estimate, let us calculate $\epsilon_\infty$ of
metallic aluminum.\cite{note_aluminum} In this case we take the 3s and
3p levels as those forming the conduction band, whereas there are 3
core levels (1s, 2s, and 2p) with $Z_{1s}=2$, $Z_{2s}=2$, and
$Z_{2p}=6$ and excitation energies $\hbar\omega_{1s}\approx 1560$~eV,
$\hbar\omega_{2s}\approx 118$~eV, and $\hbar\omega_{2p}\approx
73$~eV.\cite{Williams} For an fcc crystal with lattice constant
$a=4.05$~\AA, we obtain $(\hbar\Omega)^2\approx 83$~eV$^2$, and
therefore $\epsilon_\infty\approx
1+82\times(6/73^2+2/118^2+2/1560^2)\approx 1.11$, which compares
reasonably with the result $\epsilon_\infty\approx 1.04$ of a much
more detailed study.\cite{Smith78}

Turning to MoC, we have $(\hbar\Omega)^2\approx 71$~eV$^2$. The
relevant (not too low-lying) core orbitals $k$ for molybdenum are 4p,
4s, and 3d with $Z_k$ equal to 6, 2, and 10, respectively.  The
corresponding energies $\hbar\omega_k$ are equal to 36~eV, 63~eV, and
228~eV, respectively.\cite{Williams} For carbon, there exists only one
core orbital 1s with $Z=2$ and $\hbar\omega=284$~eV.\cite{Williams}
With these values, we find $\epsilon_\infty\approx 1.38$.

\subsection*{Robustness of the prolongation procedure}
Instead of the model
Eqs.~(\ref{eq:comprehensive1},\ref{eq:comprehensive2}) which exhibits
a cusp, let us consider the following function with the same low- and
high-frequency functional forms, which in addition contains an
intermediate frequency region from $\omega_{-}$ to $\omega_{+}$ that
smoothly connects the low- and high-frequency formulae:
\begin{equation*}
\label{eq:Grw}
\sigma^\prime(x) = \left\{
    \begin{array}{ll}
      \sigma_0(1-\mathcal{Q}^2+\mathcal{Q}^2\sqrt{x}) 
      & \text{if } 0<x<x_{-},\\
      f(x) & \text{if } x_{-}<x<x_{+},\\
      \sigma_0/(1+x^2) & \text{if } x\geq x_{+},
    \end{array} \right.
\end{equation*}
where $f(x)=a_0+a_1(x-x_-)+a_2(x-x_-)^2+a_3(x-x_-)^3$ and
$x=\omega/\Gamma$ is a dimensionless frequency. In order to proceed,
we have to fix the frequency $\omega_{-}$ where the real part of the
conductivity starts to deviate from Eq.~\eqref{eq:comprehensive1}.

The prolongation procedure proceeds as follows.  We start by guessing
the frequency scale $\Gamma$ and, with known $\Gamma$, we determine
$\sigma_0$ and ${\mathcal Q}$ from a fit of the measured data to
Eq.~\eqref{eq:comprehensive1}. The 5 remaing parameters
$a_0,a_1,a_2,a_3$ and $x_{+}$ are determined by requiring that the
function $\sigma^\prime(x)$ and its first derivative are continuous at
$x_{-}$ and $x_{+}$; furthermore we require that
$\sigma^\prime(x_{-})=\sigma^\prime(x_{+})$.  Having specified the
real part of conductivity $\sigma^\prime(\omega)$, the rest of the
prolongation procedure is the same as described in the main text: we
iterate the choice of $\Gamma$ until the Kramers-Kronig image of
$\sigma^\prime(\omega)$ reproduces the required
$\sigma^{\prime\prime}(\omega)$. 

\begin{table}[t!]
	\centering
	\begin{tabular}{ |p{2cm}||p{1cm}|p{1cm}||p{1cm}|}
		\hline
		$\hbar\omega_-$ (eV)   &  3.3  &  4.1 &  5.0  \\
		\hline
	  $g_0=Z_0\sigma_0 d$   &  1.20  &  1.20 &  1.20  \\
		$\mathcal{Q}^2$ &  0.65  &  0.65 &  0.65  \\
		%	$\Gamma/2\pi $ (THz)    &  20 &  2450 &  3280  & 2900  \\
		$\hbar\Gamma$  (eV)  & 10.4 &  10.2 &  10.1  \\
		\hline
	\end{tabular}
\caption{Parameters obtained from the fits of the normalized sheet
  conductance presented in Fig.~\ref{fig:splines} for three choices of
  the frequency $\omega_{-}$.}
	\label{tab:fit_prolong}
\end{table}

As an example, in Fig.~\ref{fig:splines} we show smoothened Lorentzian
fits of the complex conductivity from Fig.~\ref{fig:cond_exp} assuming
$\epsilon_\infty=1.4$ for three natural choices of $\omega_{-}$. As
shown in Table~\ref{tab:fit_prolong}, the smoothening procedure
essentially does not change $\sigma_0$, and $\mathcal{Q}$, while
leading to only minor changes of the parameter $\Gamma$.

\subsection*{Estimate of the 3D/2D crossover temperature}
If interaction effects dominate the quantum correction, the crossover
from 3D to 2D behaviour in a film of thickness $d$ is expected to
occur close to the temperature $k_B T_0=\hbar D/d^2$ where $D$ is the
diffusion coefficient.  Expressing $D$ in terms of the mean free path
$\ell$ we can write
$$
k_B T_0=\frac{\hbar^2}{3md^2} k_F\ell.
$$
For a film with thickness $d=5$~nm, making use of $k_F\ell=3.5\pm
0.1$ we then obtain $k_B T_0=3.4\pm 0.1$~meV which corresponds to a
crossover temperature $T_0\approx 41$~K.

\begin{acknowledgments}
We thank P.~Marko\v{s} and M.~Mo\v{s}ko for useful discussions and
A.~Weisse for sending us his numerical data. This work was supported
by the Slovak Research and Development Agency under the contract
APVV-16-0372. R.H. was supported by the Slovak Research and
Development Agency under Contract No.~APVV-15-0496.
\end{acknowledgments}


\begin{thebibliography}{99}
\bibitem{note_conductance} In the literature on transport, a different
  dimensionless conductance $\widetilde{g}=\sigma d 2\pi \hbar/e^2$ is
  often defined, which is related to our $g$ by
  $\widetilde{g}=g/(2\alpha)$, where $\alpha\approx 1/137$ is the
  fine-structure constant.
\bibitem{note_epsilon}More precisely, we are studying only the
  electronic contribution to the real part of the dielectric constant
  $\epsilon_\infty$. In principle, there exists also a similar
  contribution of the ionic degrees of freedom, which may be sizeable
  at frequencies comparable to the phonon frequencies and
  lower. However, the ellipsometric data start at $\hbar\omega\approx
  1.2$~eV, where such contribution can be neglected.
\end{thebibliography}
\end{document}